\documentclass[epsf,aps,floatfix,nofootinbib]{revtex4}
\pdfoutput=1
\usepackage{graphics}
\usepackage{graphicx}
\usepackage{color}
\usepackage{amssymb}            
\usepackage{amsmath}
\usepackage{verbatim}
\usepackage{pstricks}
\usepackage{slashed}
\usepackage{epstopdf}
\begin{document}

\setlength{\baselineskip}{0.22in}

\begin{flushright}MCTP-07-29\\
\end{flushright}
\vspace{0.2cm}

\title{Color-octet scalars at the LHC}

\author{Michael Gerbush$^{1}$, Teng Jian Khoo$^{1}$, Daniel J. Phalen$^{2}$,  Aaron Pierce$^{2}$, David Tucker-Smith$^{1}$\\
}
\vspace{0.2cm}
\affiliation{$^{1}$Department of Physics, \\
Williams College, Williamstown, MA 01267}
\vspace{0.2cm}
\affiliation{$^{2}$Michigan Center for Theoretical Physics (MCTP) \\
Department of Physics, University of Michigan, Ann Arbor, MI
48109\\}

\date{\today}

\begin{abstract}
Color-octet scalars, if present at the TeV scale, will be produced in abundance at the LHC.  We discuss in some detail the phenomenology of scalars in the $(8,2)_{1/2}$ representation, recently identified by Manohar and Wise as an addition to the standard-model Higgs sector consistent with the principle of minimal flavor violation.  Couplings of this multiplet to the Higgs lift the mass degeneracy among its states, possibly allowing for two-body decays of a heavier colored scalar to a lighter one and a gauge boson.  We perform a renormalization group analysis of these couplings and find that limits from Tevatron searches leave little room for these  decays.  This fact, and the assumption of  minimal flavor violation, lead us to study the case where the octets 
decay to the heaviest kinematically accessible fermion pairs. Focusing on pair-production events leading to $t {\overline t} t {\overline t}$, $b {\overline b} b {\overline b}$, and  $b {\overline b} t {\overline t}$ final states, we find that discovery at the LHC should be possible up to masses exceeding 1 TeV.
 \end{abstract}

\maketitle

\setcounter{equation}{0}

\section{Introduction}
In the coming year, the Large Hadron Collider (LHC) will begin its exploration of  
the high-energy frontier, and may very well uncover evidence for physics beyond the Standard Model.  
Even if LHC data indicate dramatic departures from Standard Model
predictions, determining the correct
interpretation for these deviations will likely be a complex and challenging process.
It is therefore worthwhile to consider a range of possibilities 
for what new physics might emerge, and to understand how these different possibilities would reveal
themselves experimentally.  

Exotic scalars are one possibility.  If they are to  couple to standard-model fermions  via renormalizable Yukawa couplings, the candidate representations for these scalars are actually rather limited.   
Under SU(3)$_{\rm{C}} \times$ SU(2)$_{\rm{L}}$,  the allowed representations include (8,2), (1,3), (1,2), ($\bar{3}$,3), ($\bar{3}$,2), ($\bar{3}$,1), (6,3), (6,1), (3,2),  and (1,1).   Of these, some are well-known possibilities with well-developed phenomenology, {\em e.g.} leptoquarks~\cite{LQs}.  Here, we will focus on scalars, $S$,  in the (8,2) representation, which are not as well studied.
Colored representations are of perhaps the most immediate interest, because they are the ones that will be copiously produced in the pp collisions at the LHC, assuming their masses lie near the TeV scale.  

What happens once $S$ particles are produced?
This depends in detail on their couplings to matter.  If  the $S$ fields couple to fermions very weakly or not at all, 
they  are stable for collider purposes, and the details of how they can be detected depend on how they hadronize.  In the most pessimistic case, where $S$ particles hadronize completely into neutral states, they are detectable through initial and final state radiation  in a monojet search, as discussed for long-lived gluinos in Ref.~\cite{Hewett:2004nw}.
Here we will  focus on the case where the octets {\em do} decay promptly, so that  their existence must be  inferred from their decay products.  

In thinking about these decay products we are guided by one of our main motivations for  considering color-octet scalars in the first place:   if the octets are in the the $(8,2)_{1/2}$
representation under $SU(3)_{\rm{C}} \times SU(2)_{\rm{L}} \times U(1)_{\rm{Y}}$, {\em i.e.} if they have the same electroweak  properties as the Higgs doublet, they can couple directly to quarks through renormalizable operators.  In fact, these states were singled out in  \cite{Manohar:2006ga} as {\em the} colored scalars that can couple to quarks in a way consistent with minimal flavor violation (MFV) \cite{Glashow:1976nt,MFV}.  If we insist on MFV
to suppress dangerous flavor-changing processes, then  the couplings of $S$ to fermions take the form
\begin{equation}
\label{Eqn:FermCoup}
\mathcal{L} \supset \eta_U  \bar{Q}_L y_U u_R S +\eta_D  \bar{Q}_L y_D d_R S + h.c.,
\end{equation}
where $\eta_{U,D}$ are complex numbers and $y_{U,D}$ are the Yukawa matrices.   Under the
assumption of MFV, $S$ thus couples more strongly to the heavier quarks, just like the Higgs.  
One expects $ t {\overline t}$  to be the dominant fermionic decay mode for 
the neutral components of $S$, or perhaps  $b {\overline b}$, depending on the relative sizes of the flavor-independent couplings $\eta_{U}$ and $\eta_{D}$, and on kinematic accessibility.  If kinematically allowed, the dominant fermionic decays for the charged octets would be to $t {\overline b}$ and ${\overline t} b $.   It is possible that electroweak symmetry breaking may sufficiently split the masses of the various components of $S$ so that
these decays to quarks compete with decays of a heavier $S$ component  to a lighter one and a 
gauge boson.  In Section \ref{Sec:Decays}, we discuss the likelihood of different decay modes.  In general, we find that decays to fermions are likely to dominate for all members of the $S$ multiplet.
  
Although we will focus on scalars in a $(8,2)_{1/2}$ representation, some of our analyses will
apply equally well to electroweak-singlet color octets that  couple through higher-dimension
operators to quarks.  If we impose MFV here as we did above, we have
\begin{equation}
\mathcal{L} \supset {1\over \Lambda} (\eta_U  \bar{Q}_L y_U u_R h S +\eta_D  \bar{Q}_L y_D d_R h S + h.c.).
\end{equation}
Provided that the scale $\Lambda$ for these effective operators is not too large,  $S$
could still decay promptly to quarks, preferentially to $t \overline{t}$ or $b \overline{b}$.
The phenomenology of particles in the $(8,1)_{0}$ representation was explored  in a supersymmetric context in \cite{Yanou}.

The outline for the rest of the paper is as follows.  In next section,  we discuss contexts in which octet scalars might arise.  In Section \ref{Sec:MWScalar}  we study the most general renormalizable scalar potential involving the Higgs doublet and an $S$ multiplet in the $({ 8},{ 2})_{1/2}$ representation.    We show that  two-body decays of the octets involving gauge bosons are somewhat disfavored by an analysis of the renormalization group equations for these couplings, combined with limits from Tevatron searches.  Motivated by this result, we focus
in Sections IV-VI on $S$ decays to quarks, $t {\overline t}$, $b {\overline b}$,   and $t {\overline b}$  in particular.  In Section IV we discuss the challenges of reconstructing top quarks in $S$ decays and present our strategy, which is to identify fat composite jets with invariant masses near the top mass as top candidates.   In Section V we apply this method to $S$ pair-production, in which case  the cross section depends on the mass alone, as the relevant coupling is the SU(3)$_{\rm{C}}$ gauge coupling.  We also briefly consider the more parameter-dependent case of single $S$ production, which appears to be more experimentally challenging.  Finally, in Section VI we consider how one might hope to probe the couplings of $S$ to the Higgs.  

\section{Motivation}\label{Sec:Motivation}
As we are chiefly interested in the collider phenomenology of color-octet scalars, we will
for the most part set aside questions about where they come from and why they should exist.  
Here we simply mention two possibilities for how they could fit into a larger theory.  

In theories of compositeness, it is possible that color-octet states could emerge if some
or all of the fermions transforming under the confining gauge group are colored.    If some
fermions transform as fundamentals of SU(3)$_{\rm{C}}$ and others as anti-fundamentals, it is perfectly 
reasonable to imagine that color-singlet bound states, appropriate to serve as the Higgs doublet, could
be accompanied by color-octet bound states, given the decomposition $3 \otimes {\overline 3} = 8+1$.
Assuming that (anti-)fundamental representations are likely to emerge from a high-scale theory, this simple group theory relation is already motivation for octets.

To make this point slightly more explicit, consider the following cartoon of a composite Higgs model.  Assume  left-handed fermions in the following representations under
SU(N)$\times$SU(3)$_{\rm{C}} \times $SU(2)$_{\rm{L}} \times$ U(1)$_{\rm{Y}}$, where SU(N) is the confining gauge group:
\begin{equation}
\psi \sim (N, { 3},{ 2})_{1/2} \quad \quad \quad \quad \psi^c \sim ({\overline { N}}, {\overline { 3}},2)_{-1/2} 
\end{equation}
\begin{equation}
\chi \sim ({ N},{ 3},{ 1})_0 \quad \quad \quad \quad \chi^c \sim ({\overline { N}},{\overline { 3}},1)_0. 
\end{equation}
These femions are in vector-like representations, but we will imagine that their masses lie well
below the SU(N) scale.

Setting the gauge couplings of the Standard Model to zero, this theory exhibits an SU(9)$\times$SU(9) flavor symmetry.
If   the strong SU(N) dynamics break this flavor symmetry to its diagonal subgroup with the equal condensates
\begin{equation}
\langle \psi \psi^c \rangle =  \langle \chi \chi^c  \rangle,
\end{equation}  
then in the limit where the Standard Model gauge couplings are turned off, there are 80 Goldstone bosons, including real scalars transforming as  $({ 8},{ 1})_0 \oplus ({ 8},{ 1})_0 \oplus  ({ 8},{ 3})_0 \oplus  ({ 1},{ 3})_0 \oplus  ({ 1},{ 1})_0$ under the Standard Model gauge group, and complex scalars transforming as $({ 8},{ 2})_{1/2} \oplus ({ 1},{ 2})_{1/2} $. All of these particles receive quadratically-divergent masses from gauge loops except for the singlet, but if  $\psi$ and $\chi$ have different masses, that symmetry breaking would presumably induce a mass for the singlet.  

The   $({ 1},{ 2})_{1/2}$ state can be identified as the Higgs doublet of the Standard Model, although
we have not specified here an origin for its quartic coupling, and  have made no attempt to provide a mechanism to achieve natural electroweak symmetry breaking.  Among the
many additional states are those in the   $({ 8},{ 2})_{ 1/2}$ multiplet.   In what follows, we will focus on the phenomenology of this representation, although, as this
example makes clear, if that multiplet exists it is plausible that others will be present as well.  

A color-octet scalar can also emerge as an extra component of the gluon field in an
extra dimensional model with bulk gauge fields (or, more precisely, as one of the low-lying Kaluza-Klein excitations of that field)~\cite{Bogdan}.   Of course, in this case the scalar octet will not
have electroweak charge, but decays to quarks could still be induced by brane-localized couplings involving the Higgs.
If these brane-localized couplings have hierarchies that are similar to the Yukawa hierarchies, the octet will decay preferentially to third-generation quarks.

\section{Color Octet Doublets and their Decays}\label{Sec:MWScalar}

We now consider the phenomenology of a color-octet scalar $S$ in the $(8,2)_{+1/2}$ representation under SU(3)$_{\rm{C}} \times$ SU(2)$_{\rm{L}} \times$ U(1)$_{\rm{Y}}$.
The couplings of $S$ to fermions shown in Eqn.~\ref{Eqn:FermCoup}
preserve MFV.  The Standard Model U(3)$^{5}$ flavor symmetry is broken only by the Yukawa couplings and by the $S$ couplings, which are themselves proportional to the Yukawa couplings.  Thus, the Yukawa couplings are the only spurions giving rise to flavor violation in this model.

Following the notation of \cite{Manohar:2006ga},
the most general renormalizable scalar potential for this model is: 
\begin{eqnarray}
V&=&\lambda(H^{\dag i} H_i - \frac{v^2}{2})^2 + 2 m_S^2 Tr\,S^{\dag i} S_i +\lambda_1 H^{\dag i} H_i Tr\, S^{\dag j} S_j +\lambda_2 H^{\dag i} H_j Tr\, S^{\dag j} S_i \nonumber \\
&&+(\lambda_3 H^{\dag i} H^{\dag j} Tr\, S_i S_j +\lambda_4 H^{\dag i}  Tr\, S^{\dag j} S_j S_i +\lambda_5 H^{\dag i} Tr \, S^{\dag j} S_i S_j + h.c.) \nonumber \\
&&+\lambda_6 Tr\,  S^{\dag i} S_i S^{\dag j} S_j +\lambda_7 Tr\, S^{\dag i} S_j S^{\dag j} S_i +\lambda_8 Tr\, S^{\dag i} S_i Tr\, S^{\dag j} S_j +\lambda_9 Tr\,S^{\dag i} S_j Tr\, S^{\dag j} S_i \nonumber \\
&&+\lambda_{10} Tr \, S_i S_j S^{\dag i} S^{\dag j} +\lambda_{11} Tr\, S_i S_j S^{\dag j} S^{\dag i}
\end{eqnarray}
where i, j are SU(2)$_{\rm{L}}$ indices and the trace is over the SU(3)$_{\rm{C}}$ indices.  Note that $\lambda_{4}$ and $\lambda_{5}$ may be forbidden by an $S \rightarrow -S$ symmetry.  This symmetry is, however, broken
by the couplings of $S$ to fermions, and so  even if $\lambda_{4}$ and $\lambda_{5}$ vanish at tree-level, they will be generated radiatively by $\eta_{U}$ and $\eta_{D}$.  If  $\eta_{U,D}$ and  $\lambda_{4,5}$ {\em all} vanish, then the $S$ field becomes exactly stable.  In this case, very stringent limits from heavy elements searches apply.   For the remainder of this note, we will assume that the $S \rightarrow -S$ symmetry is broken strongly enough to avoid all cosmological problems, including those that might arise from late-time decays \cite{GluinoCosmo}.

Cross couplings between the $S$ and $H$ fields can significantly modify the properties of the Higgs boson.  Previous work has noted that these couplings can lead to a substantial effect on, e.g., the $gg \rightarrow h$ production rate \cite{Manohar:2006ga}.  We will revisit related issues in Section \ref{Sec:HiggsCouplings}, but first we discuss observation of $S$ fields at hadron colliders.

\subsection{Decay Modes}\label{Sec:Decays}
As a prelude to our discussion, we consider which decay modes of $S$  are likely to dominate.  Working under the assumption of minimal flavor violation, the lightest $S$ state will decay to the heaviest fermion states that are kinematically accessible.  How the other states in the $S$ multiplet decay is slightly more subtle.  The answer depends on the couplings of  $S$ to fermions, as well as the splittings within the $S$ multiplet.
The $S$ doublet will be split by electroweak symmetry breaking: 
\begin{eqnarray}
m_{S^+}^2 &=& m_S^2 + \frac{v^2}{4}\lambda_1 \nonumber \\
m_{S^R}^2 &=& m_S^2 + \frac{v^2}{4}(\lambda_1+\lambda_2+2\lambda_3) = m_{S^+}^2 + \frac{v^2}{4}(\lambda_2+2\lambda_3)  \nonumber \\
m_{S^I}^2 &=& m_S^2 + \frac{v^2}{4}(\lambda_1+\lambda_2-2\lambda_3) = m_{S^+}^2 + \frac{v^2}{4}(\lambda_2-2\lambda_3) = m_{S^R}^2 - v^2\lambda_3,
\end{eqnarray}
where $S^+$, $S^R$, and $S^I$ are the charged, neutral scalar, and neutral psudoscalar components of $S$.
If the splittings are sufficiently large, then two-body decays of the type $S^{+} \rightarrow S^{0} W$ and $S_{I} \rightarrow Z^{0} S_{R}$ are allowed, and  production of the heavier states in the multiplet  leads to cascade events.  However, as we now discuss, this scenario does not appear to be likely. 

To determine the size of the mass splittings induced by the above couplings, we need to known what to expect for both $m_{S}^{2}$ and the $\lambda_{i}$.  We can get an estimate of the current bound on $m_S^{2}$ by considering searches at the Tevatron for final states with multiple heavy jets.  Such a search was recently conducted by CDF, and was used to place limits on the process $gg \rightarrow A^{0} b \bar{b}$, where $A^{0}$ is the pseudoscalar Higgs of the MSSM~\cite{CDFNote}.  Roughly, this result indicates that fields that decay to 4$b$ with $\sigma \times$ Br($\rightarrow 4b$) $\approx$  1 pb are likely to be excluded.  We generated events of the type $gg \rightarrow SS \rightarrow 4b$ using the {\tt UsrMod} functionality of Madgraph  {\tt v.4.1.30} \cite{MadGraph}, using the CTEQ5L parton distribution function.  We applied a K factor of 1.5, in analogy to gluino production at the Tevatron \cite{Kfactors}.  Events were subsequently showered using Pythia 6.4.11\cite{Pythia}, and then run through the PGS detector simulation (v4)\cite{PGS}\footnote{We modified the default b-tagging parameterization of  PGS.  The standard PGS distribution determines whether a $b$-jet is tagging by looking at the Monte Carlo ``truth'' information for tracks within 20 degrees of a reconstructed jet center, and then applies a $(p_{T}, \eta)$ dependent efficiency.  This approach is only reasonable if the jet size is close to 20 degrees.  Since we look at a variety of cone sizes, we modify the PGS code to look for "true" b quarks within a cone whose size depends on the cone size used in jet reconstruction.}. 

 Placing cuts similar to those in \cite{CDFNote} on the signal, we find that current limits require $m_S \gtrsim 200$ GeV if $S$ decays dominantly to $b$ quarks.    We should note that even if $S$ is too light to decay to $t{\overline t}$, there is no guarantee that $b {\overline b}$ will be the dominant decay mode.   If   $\eta_{D}/\eta_{U}$ is sufficiently small, then $S$ may dominantly decay to charm quarks with a resulting decrease in the efficiency of the quoted search.  In this case, the bound could be substantially degraded.  
 
\begin{figure}[ht]
\begin{minipage}[b]{0.5\linewidth}
\centering
\scalebox{0.5}{ \includegraphics{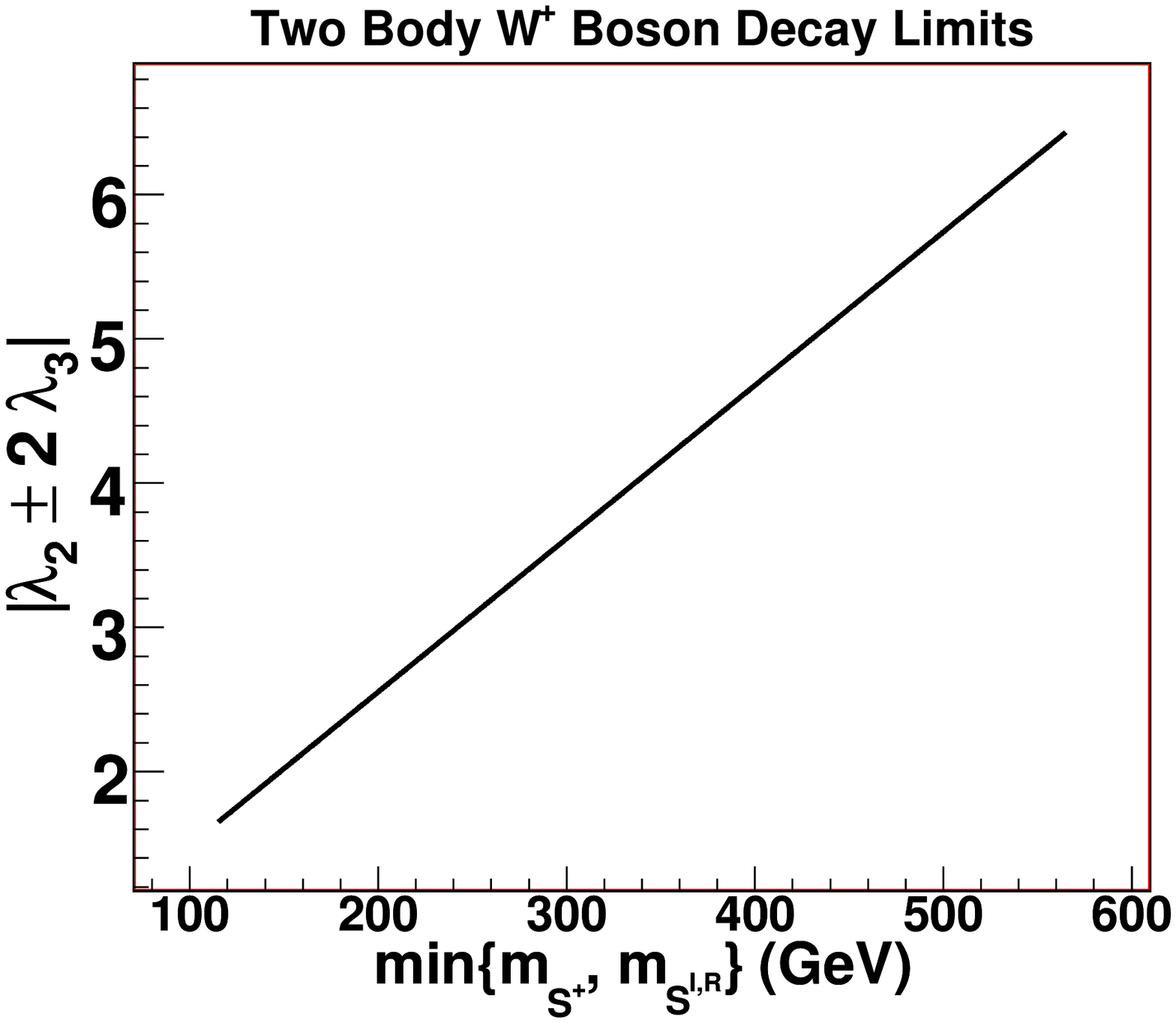}}
\caption{The minimum value of $|\lambda_2 \pm 2\lambda_3|$ that allows for two-body decays  involving $W^+$.  The positive sign corresponds to $S^R \to W^+ S^-$ or $S^- \to W^-  S^R $, and negative corresponds to  $S^I \to W^+ S^-$ or  $S^- \to W^- S^I$. \label{fig:Wdecaylimits}}
\label{Fig:Wbounds}
\end{minipage}
\hspace{0.5cm}
\begin{minipage}[b]{0.5\linewidth}

\centering
 \scalebox{0.5}{ \includegraphics{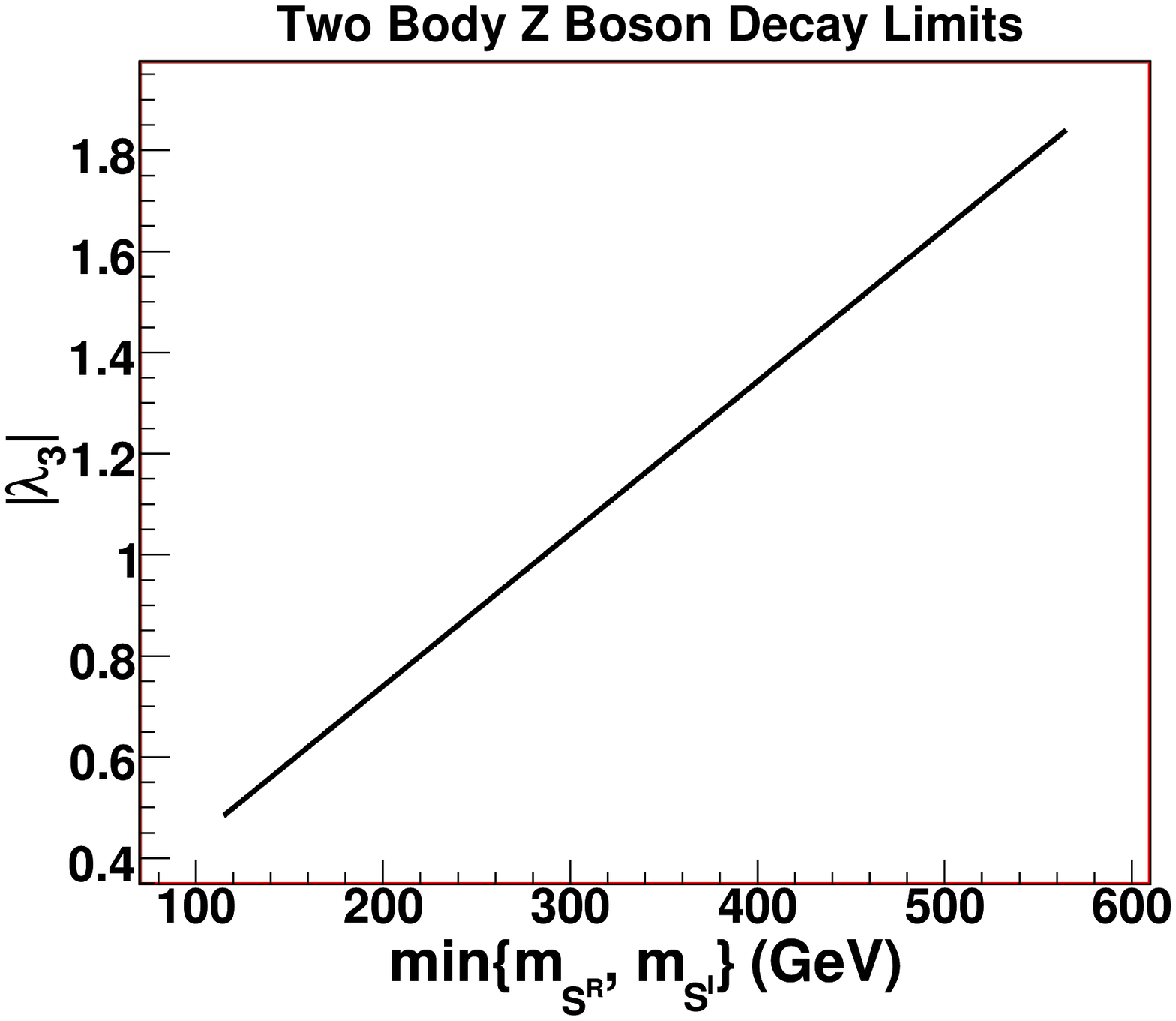}}
\caption{The minimum value of $|\lambda_3|$ that allows for $S^I \to Z S^R$ or $S^R \to Z S^I$. \label{fig:Zdecaylimits}}
\label{Fig:Zbounds}
\end{minipage}
\end{figure}

In what follows we will assume that the decays to $b$ quarks dominate, and take the bound to be 200 GeV. With this bound in place, Figures \ref{fig:Wdecaylimits} and \ref{fig:Zdecaylimits} indicate that on-shell gauge boson decays are only possible for large $\lambda_{i}$.  With such large couplings, the Higgs quartic coupling tends to quickly go non-perturbative (see Appendix for RGEs), indicating the need for new physics at a relatively low scale.  First, suppose we impose the very modest requirement that the Higgs quartic coupling remain preturbative up to the 10 TeV scale.  In this case, for a Higgs boson mass of 120 GeV, one can accomodate $\lambda_{1} \lesssim 1.3$ 
or $\sqrt{\lambda_{2}^{2} + \lambda_{3}^{2}} \lesssim 2.2$, in both cases assuming all the other $\lambda_{i}$ are negligible. 
 For a larger Higgs boson mass, of say 200 GeV,  the bounds are even stronger, $\lambda_1 \lesssim 1.2$, and  $\sqrt{\lambda_{2}^{2} + \lambda_{3}^{2}} \lesssim 1.8$.  From Figs. \ref{fig:Wdecaylimits} and \ref{fig:Zdecaylimits}, we can conclude that even the modest requirement of having no Landau pole below 10 TeV significantly limits the space where gauge boson cascades are allowed.  If  we instead impose the absence of a Landau pole all the way up to $10^{10}$ GeV, the limits tighten considerably, forcing $\sqrt{\lambda_{2}^{2} + \lambda_{3}^{2}} \lesssim 0.9$ for a 120 GeV Higgs boson.

Finally, we note that large $\lambda_{2,3}$ together with small $m_{S}$ can lead to large contributions to precision electroweak observables  \cite{Manohar:2006ga}.    For somewhat light octets, $m_{S} \approx 300$ GeV, the constraints on $\delta \rho$ require $(\lambda_{2}^{2} -4 \lambda_{3}^{2})\lesssim 1$, again disfavoring two-body decays to gauge bosons.

The above considerations indicate that all members of the $S$ multiplet are likely to decay to the heaviest available fermions.   It should be noted, however, that the width to quarks does depend on the unknown paramters $\eta_{U}$ and $\eta_{D}$.  If these parameters are sufficiently small, then three-body gauge boson mediated decays can still compete.  As an example, for $m_{S^+} = 830$ GeV and $m_{S^0} = 800$ GeV, the three-body decay mediated by W bosons can compete if $\eta_U \lesssim 10^{-5}$ and $\eta_D \lesssim 10^{-3}$.  Finally, if the $\eta_{i}$ are very small, {\it and} the splittings between the charged and neutral states are very small ($\lesssim$ GeV), then there is a possibility of long-lived charged tracks with displaced vertices (or even highly ionizing tracks that range through the detector).  For a review of some of the relevant phenomenology in this case, see \cite{Kraan}.    An exotic in the (8,1)$_0$ representation  couples to the Standard Model fermions only through higher dimension operators, suppressed  by a high scale, $M$. For sufficiently high $M$, this provides a natural explanation for a long-lived exotic.
In the remainder of the paper, we will assume that the $\eta_{i}$ are sufficiently large the the decay to heavy  fermions will dominate.  

\section{Reconstruction of Top Quarks in S decays}
Since  $S$ is charged under SU(3)$_{\rm{C}}$, one would expect the LHC discovery reach to be impressive, a TeV or greater. If $S$ is very heavy and decays to $t \overline{t}$, reconstructing these top quarks finding may present challenges due to the kinematics of the events. Because the  top quarks are boosted in the lab frame,  the angle between their decay products -- the W boson and b-quark -- tend to be small.  Thus, there are two potentially useful methods to reconstruct the top quark: one can attempt to reconstruct a  W boson from its decay products and then combine it with a b-jet as usual, or one can look directly for jets with invariant mass near the top quark mass \cite{ATLASTDR, Butterworth:2002tt, DTSWS,Holdom}.  If the W boson decays hadronically, then there are three jets very close and possibly overlapping in the detector.  It is possible that a jet algorithm might reconstruct this as a single jet, in which case its mass should correspond to that of the top.  

In the case where the jets are resolved by the jet finding algorithm,  a conventional two step reconstruction may be attempted.   If one is tagged as a b-quark, then the other two jets (or one if the W boson jets are overlapping) can be used reconstruct the W. This W can then be subsequently combined with the b-jet to reconstruct a top.  Particularly in the highly populated final states that we consider here, this method can potentially suffer from large combinatoric backgrounds and high fake rates.  Light jets from initial and final state radiation can fake $W$'s, leading to a pollution of the signal.  Alternately, in events with exactly one lepton, the W boson may be reconstructed by  combining the lepton with the missing energy.  Degeneracies in $\eta$ for the $\slashed{p}_T$ reconstruction are broken by looking for what best will reconstruct the top quark.  This method was recently used in reconstructing  single heavy resonances decaying to top quarks~\cite{Barger:2006hm}.

\subsection{A Top Quark Finding Strategy}

We find that an efficient way of searching for top quarks is to look for fat ``top jets.''  One might be tempted simply to analyze events with a large cone size, hoping that all top decay products would fall into the cone.   However,  these events are so highly populated (including the decay products of up to four top quarks,  ISR and FSR) that a large cone size is apt to suck up extra soft things from far away into a top jet, thus moving the jet mass out of the window for top-quarks.  To combat this, a two step process is used.  First, when reconstructing jets in the event a small $k_{T}$ cone size of $\Delta R = 0.3$ is initally used.  Second,  after the jet reconstruction is completed, there is a jet combination phase.  The distance $\Delta R_{jj}$ between all pairs of jets is calculated.  If $\Delta R_{jj_{min}} < \Delta R_{combine}$, then those two jets are merged.  This process is repeated until all jets are at least $\Delta R_{combine}$ apart.\footnote{Note that by using the $k_T$ algorithm initially, this approach preserves nice theoretical properties such  infrared-safety.}

This $\Delta R_{combine}$ is chosen based on the expected boost of the top quark, so that all decay products will be captured.  It will be somewhat larger than simple kinematic calculations would expect, since the $k_T$-jet algorithm can spread out nearby jets dependent on how nearby soft tracks are clustered.  Composite jets with an invariant mass between 125 GeV and 225 GeV are called top jets.

We may optionally augment this procedure to account for leptonically decaying W bosons.  In this case, we reconstruct W bosons leptonically, then merge these with b-jets to find tops.  After this, we use the jet merging method described above to reconstruct any remaining tops that decay entirely to jets.  Whether including these leptonically decaying top candidates is advantageous depends on the particular search in question.  In the discussion below we will indicate which analyses include them and which do not.  

In passing we note that the determination of the invariant mass of the jet will be potentially sensitive to complications such as the underlying event and multiple interactions.  We will assume that a subtraction of these superfluous energy depositions can be done efficiently.  This procedure should be able to be callibrated on other samples (e.g., t-tbar production).

With this procedure for finding top jets in hand, we now move to analyze the prospects for finding $S$ pair production at the LHC.

\section{Finding the S fields at Colliders}

Pairs of $S$ fields may be produced copiously, leading to final states with up to four top quarks.   We can attempt to reconstruct top jets as above, and then take pairings to look for resonances.  Though there will be a combinatoric background from incorrect pairings of  the top jets, we will show that it is still possible to find the $S$ resonance.

\subsection{Neutral Scalar Pairs}

As a benchmark, we will consider the production of a pair of neutral $S$ states with mass of 1 TeV.  We take the psuedoscalar and scalar pieces to be degenerate, for it is unlikely that the electroweak splitting will be appreciable (i.e. larger than our reconstruction resolution) at that these masses.  We will  first
assume that $\eta_{U}$ is not much less than $\eta_{D}$, so that decays to top quarks dominate, and then discuss the case where  decays to bottom quarks dominate.

At the LHC, the production cross section for pair production of the scalar and pseudoscalar for a mass of 1 TeV is calculated to be 176 fb, where we have included a K-factor of 2 in analogy with that found for a gluino of similar mass at the LHC \cite{Kfactors}.  At this mass, the top quark boost in the rest frame of the $S$ particle will be $\sim 3$.  To eliminate backgrounds from $W $+ jets and pure QCD, we perform a preselection. After these cuts, the largest backgrounds are expected to come from $t\bar{t} +$ jets production.  We generate a $t \bar{t}$+jets sample using Alpgen  2.12 \cite{Mangano:2002ea}, with showering and hadronization performed by Pythia.   We use the jet-parton matching procedure built into Alpgen to obtain $t\bar{t} +0,1$ jets exclusive samples and a $t\bar{t} +2$ jet inclusive sample.   To avoid having to generate an enormous event sample, we impose a 1-TeV generator-level cut 
on the scalar sum of the $p_T$'s of the final-state partons (so, the six decay products of the tops, and any additional jets).  Given the much more stringent cuts applied at the analysis level, this generator-level cut is not problematic.  To estimate the luminosities of our background samples we multiply the cross sections given by Alpgen and Pythia by a $K$ factor
determined by comparing the Alpgen $t {\overline t}$ cross section in the absence of cuts to the NLO result, 830 pb \cite{Bonciani:1998vc}.  We generated roughly 100 fb$^{-1}$ 
of integrated luminosity, and have rescaled the background levels appropriately in the results presented below.

    As before, signal events are generated using the {\tt UserMod} interface to MadGraph \cite{MadGraph}, and subsequently showered in Pythia.  Finally, the events are run through the PGS detetor simulation package.  Events are generated with a renormalization scale $ \mu^2 = (M_{FS} + \sum_j{ p_T^2})$, where $j$ runs over all particles, and $M_{FS}$ is the maximum mass in the final state.

\begin{figure}
\begin{center}
\scalebox{0.6}{ \includegraphics{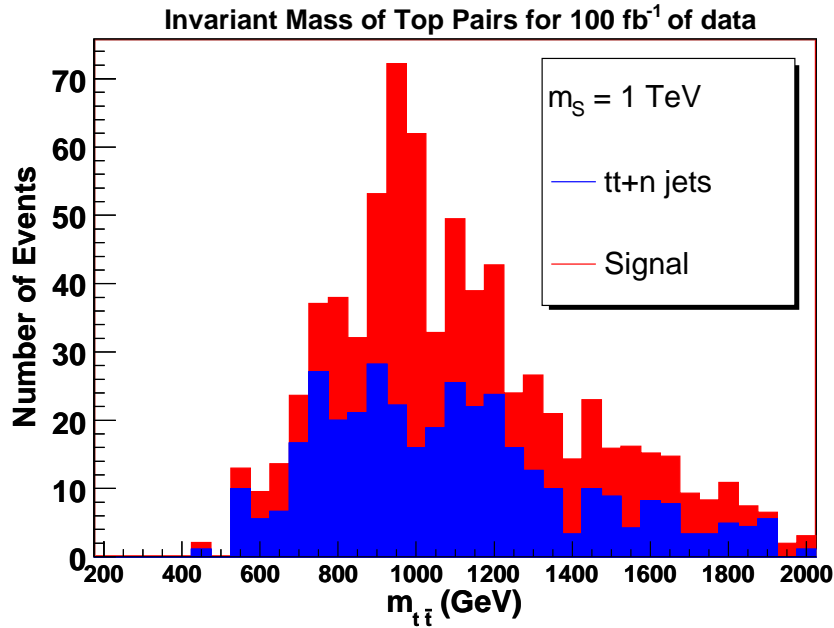}}
\caption{ Invariant mass distributions from reconstructed top quark pairs, for signal and $t{\overline t}$ background.  The signal is stacked on the background.
 Looking in the mass window  $1 \pm 0.1$ TeV gives a signal significance of over 5$\sigma$.  }
 \label{fig:mx1000bkgsig}
\end{center}
\end{figure}

The following preselection cuts were imposed:

\begin{itemize}
\item
Two b-jets with $p_T>50$ GeV.
\item
At least one electron or muon with $p_T>20$ GeV.
\item
Transverse sphericity $> 0.15$.
\item
Four jets with $p_T > 100$ GeV.
\item
$M_{eff}> 1500$ GeV.
\end{itemize}
Where $M_{eff}$ is defined as the scalar sum of all objects (including missing $p_{T}$) reconstructed in the event.   Jets are reconstructed using the $k_T$ algorithm with a cone size of $\Delta R=0.3$, and tops are reconstructed with $\Delta R_{combine}=1.1$ as described in the previous section.  In events where two or more top jets are successfully reconstructed, all top quarks are taken pairwise, and the invariant mass of each pair is plotted.  The resulting invariant mass peak is shown in Figure~\ref{fig:mx1000bkgsig}.  A signal peak is clearly visible above the background.  More sophisticated tools could likely improve the signal to background ratio and sharpen the mass peak.  For smaller masses, the production cross section will increase, but it may be necessary to adapt the cuts.  In particular, it may be advantageous to loosen the $M_{eff}$ cut.

Note that this plot was generated without attempting to reconstruct top quarks leptonically.  While this decreases the number of signal events, we found that this strategy maximizes the significance $\frac{S}{\sqrt{B}}$, as it also decreases the number of spurious pairings from the $t \bar{t}$  background.

\begin{figure}[h] 
\centering
\scalebox{1.0}{\input{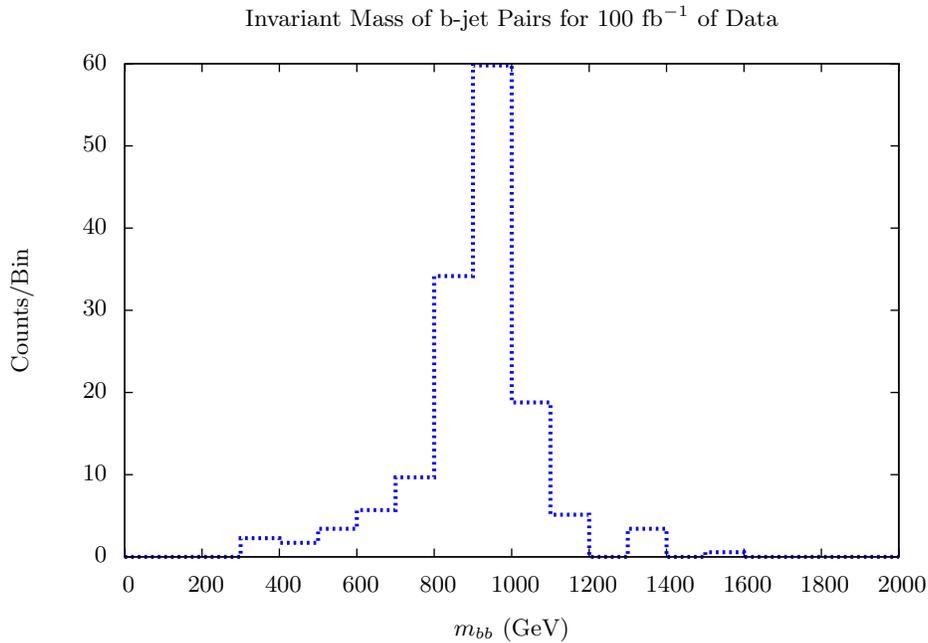}}
\caption{Invariant-mass distribution for pairs of b-tagged jets, for the case where the neutral $S$ states decay dominantly to $b {\overline b}$.}
\label{fig:bb}
\end{figure}

We now turn our attention to the scenario in which the neutral $S$ states decay dominantly to $b {\overline b}$ instead of $t {\overline t}$.  We will continue to take the mass of the neutral $S$ states to be 1 TeV. If decays to  $b {\overline b}$  dominate, no leptons appear in the final state, but we
nevertheless find that an efficient suppression of standard-model backgrounds can still be achieved using the following cuts:
\begin{itemize}
\item
Exactly four jets with $p_T > 200$ GeV, all of them b-tagged.
\item
$M_{eff} > 1800$ GeV.
\end{itemize}
For events passing these cuts, the four b-tagged jets are paired up so that the difference between
the two dijet invariant masses is minimized. If these two invariant masses are within 200 GeV of each other, both values are kept, and  otherwise the event is discarded.  

To analyze the octet signal, we generate $\sim$31k events, corresponding to $\sim$175 fb$^{-1}$.  Of
these, 141 pass the b-jet multiplicity and $M_{eff}$ cuts, and 127 satisfy the dijet invariant mass criterion.
In figure~\ref{fig:bb} we show the invariant mass distribution for these signal events after rescaling the counts  to correspond to an integrated luminosity of 100 fb$^{-1}$.   

Using Alpgen, we have considered backgrounds
to this signal from multijet production, $b {\overline b}$+jets, $b {\overline b}b {\overline b}$, and $t {\overline t}$+jets.   To estimate the multijet background we first determine
the rate for events passing the jet multiplicity cut -- without a b-tag requirement -- and the $M_{eff}$ cut.
We obtain 1,358 events passing these cuts from a sample corresponding to an integrated luminosity of 148 pb$^{-1}$ (after including a $K$ factor of 2), and find that 2.5\% of the energetic jets are identified as b-tagged
by PGS.  Estimating that the fraction of events that pass the b-jet multiplicity cut will then be $\sim(0.025)^4$,
we get $\sim 0.4$ for the expected number of events after 100 fb$^{-1}$, before the invariant-mass cut.

For the $b {\overline b}$+jets background we obtain 841 events passing the jet-multiplicity cut from a  2-fb$^{-1}$ sample (again taking $K=2$), 99 of which have two b-tagged jets.
Taking the 2.5\% fake-rate from the multijet sample, we estimate the probability for the events
passing the jet-multiplicity cut to have four b-jets to be $\sim (99/481) \times (0.025)^2$, leading to
an estimate of  $\sim 3$ events per 100 fb$^{-1}$ before the invariant-mass cut.  By studying the sample
obtained without imposing the b-tag requirement, we estimate that the invariant-mass cut reduces the background by a further factor of two or so.  We find that the backgrounds from $b {\overline b}b {\overline b}$ and $t {\overline t}$+jets are considerably smaller, with a combined expected rate of   $\sim 0.6$ events per 100 fb$^{-1}$ before the invariant-mass cut.  
%

These background estimates suffer from large QCD uncertainties and are furthermore
sensitive to b-tag efficiencies and fake-rates, approximated here using the default settings
in PGS.  Nevertheless, given that our estimates lead to a signal to background ratio well over
10, it is at least plausible that an octet signal associated with decays to $b {\overline b}$ would be
observable using an analysis along the lines of the one presented here.

\subsection{Charged Scalar Pairs}
Next, we consider whether decays of the type $S^{+} S^{-} \rightarrow t \bar{t} b \bar{b}$ can be reconstructed.  The production rate for $gg \to S^+ S^-$ pair production of the charged scalars for a mass of 1 TeV will be identical to the combined rate for the scalar plus pseduoscalar, so we again take 176 fb for the cross section.   After preselection, the largest backgrounds are again expected to come from $t\bar{t} +$ jets production, which we simulate with Alpgen 2.12.  In this case, the following preselection cuts were imposed:

\begin{itemize}
\item
One b-jet with $p_T>200$ GeV, two b-jets with $p_T > 100$.
\item
At least one electron or muon with $p_T>20$ GeV.
\item
Transverse sphericity $> 0.15$.
\item
Four Jets with $p_T > 100$ GeV.
\item
$M_{eff} > 1800$ GeV.
\end{itemize}

\begin{figure}
\begin{center}
\scalebox{0.6}{\includegraphics{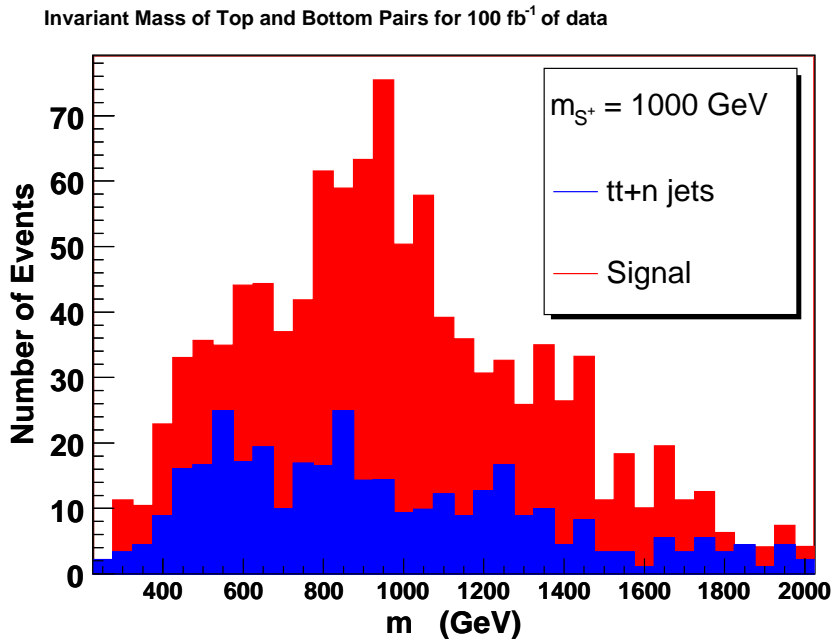}}
\caption{ Invariant mass distributions from $b$-jet -- reconstructed top quark pairs, for charged-octet signal and $t{\overline t}$ background.  The signal is stacked on the background.
 Looking in the mass window $1 \pm 0.1$ TeV gives a signal significance of over 5$\sigma$.}
\label{fig:chargedmx1000bkgsig}
\end{center}
\end{figure}

Jets are reconstructed using the same procedure used in  the neutral scalars, using the $k_T$ algorithm with a cone size of $\Delta R=0.3$, and tops are reconstructed with $\Delta R_{combine}=1.1$.  Charged $S$ resonances are then sought by combining any top quarks with any bottom quarks, and then ploting their invariant masses.  This results in the peak shown in Figure~\ref{fig:chargedmx1000bkgsig}.

Again, the signal is clearly visible above the background.  In this case, the width of the peak is greater, due in part to incorrect pairings of top candidates with b jets.  For this analysis we find that it is beneficial to include the leptonically reconstructed top candidates.

\subsection{Single Scalar production}

Although parameter-dependent, it is possible  $S$ resonances will be produced singly via gluon-guon fusion~\cite{GreshamWise}.  This process proceeds via a loop diagram, and depends on precisely  the couplings that break the $S \rightarrow -S$ symmetry.  In particular, if $\eta_{U,D} $ and $\lambda_{4,5}$ are all zero, then the cross section vanishes.  For low values of $m_S$, a large $\eta_{U}$ is in tension with the constraint from $Z\rightarrow b \bar{b}$ derived in \cite{GreshamWise},  but $\eta_{U} \approx 1$ is allowed at the two-sigma level for $m_{S} \gtrsim 500$ GeV. If $\lambda_{4,5}$ is zero, then the production in this channel is identical to gluon-gluon fusion to a Higgs boson, modulo a simple rescaling by $\eta_{U}^{2}$ and a modified color factor.  In the following, we assume $\eta_{U} =1$ and $\lambda_{4,5} =0$.  In the narrow width approximation, we then find $\sigma(gg\to S_R) = \frac{5}{12}\sigma(gg \to h)$.
 
Large $\eta_U$ means that $BR(S_R \to t\bar{t}) \sim 1$ as long as this state is kinematically accessible.  For masses below 2 $m_{t}$, decays to $b$ quarks are likely, with a staggering QCD background.  
 
The ATLAS collaboration has done a study on searching for resonances in the $t \bar{t}$ channel in the context of searches for the $A^{0}$ of the MSSM at low $\tan \beta$.  We can use this related study to determine the prospects for searching for the $S$ in this channel.   For $m_S =370$ GeV, we find $\sigma(gg\to S_R) \approx 1.7$ pb.  In Table 19-39 of  the ATLAS TDR \cite{ATLASTDR}, the ATLAS study used a benchmark point with a cross section $\sigma_{ggA^{0}}$ = 8.3 pb at this mass.  Rescaling the signal, we find that  even with $\eta_{U} =1$, 30 fb$^{-1}$ of data will only provide approximately a 1.7$\sigma$ excess in this channel.
 
\section{Probing Couplings of the $S$ to the Higgs boson}\label{Sec:HiggsCouplings}
Once discovery of an exotic $S$ is established, it is natural to ask wether it couples to the Higgs boson.  After all, the coupling $|h|^{2} |X|^2$ is allowed, regardless of the quantum numbers of the $X$ field.  Our $S$ field is no exception.  

There are several ways in which one might gain indirect access to couplings between  $S$ and the Higgs boson.  The rate for $gg \rightarrow h$ production may be modified if the $S$ couples to the Higgs boson.\cite{Manohar:2006ga}.   Also, the the presence of splittings between the members of the $S$ multiplet is due to the $S$ couplings to the Higgs bosons.  So, if finite splittings could be observed (perhaps through cascade decays via gauge decays to the lightest $S$ field), then one might be convinced that there is an interaction between  $S$ and the Higgs.  There are two smoking guns for establishing such a coupling. First, one might look for exotic Higgs decays of the form $h \rightarrow S S$.  Second, one could look directly for an $SSh$ final state, where the Higgs is radiated from an $S$ field.  

The current bound on the $S$ mass, $m_{S} \gtrsim 200$, tells us that Higgs boson decays to the $S$ field will require a heavy Higgs boson.  Given the indications from precision electroweak physics for a light Higgs boson this possiblity is somewhat disfavored.  However, one should note that the $S$ field itself can contribute to precision electroweak observables, possibly allowing for a heavier Higgs boson, along the lines of \cite{Lawrence}.    A heavy Higgs boson already has a substantial branching fraction to on-shell $W$ bosons, rendering  $S$ pairs an unlikely dominant decay channel.  However, if the $S-S-h$ coupling is large enough, then it is possible that this decay mode could at least compete with the $WW$ decay mode.   For example, for $m_{X} =200$ GeV, and $\lambda_{SSh}= v$,  the branching ratio can reach 25\% for a 500 GeV Higgs boson.  Discovery of such Higgs boson decays would be challenging, as the final state will again be four heavy fermions, but in this case a large $M_{eff}$ cut will not be possible.  If the $S \rightarrow -S$ approximate symmetry is very good (including very small $\eta_{i}$), then observation should be much easier.  In this case, the $S$ fields can be quasi-stable, and depending on the details of hadronization, they might appear in the detector as highly ionizing, long-lived charged tracks. Then the distinctive decay  $h \rightarrow  SS$ might be simple to see above Standard Model backgrounds.  However, current searches for such highly-ionizing particles at the DZero experiment constrain particles with an $S$-like production cross section to have masses  above 310 GeV \cite{DZeroNote}.  Then, even a 700 GeV Higgs would only have a $5\%$ branching fraction to such states, even with $\lambda_{SSh} =v$.  Whether or not the Higgs boson mass peak could be reconstructed in the final state above the pair produced $S$ continuum is an interesting question.  The answer depends in detail upon how well the mass of the $S$ particles could be measured, from a combination of $dE/dx$ measurements and time of flight considerations.

\begin{figure}
\begin{center}
\label{Fig:XXh}
\scalebox{0.6}{ \includegraphics{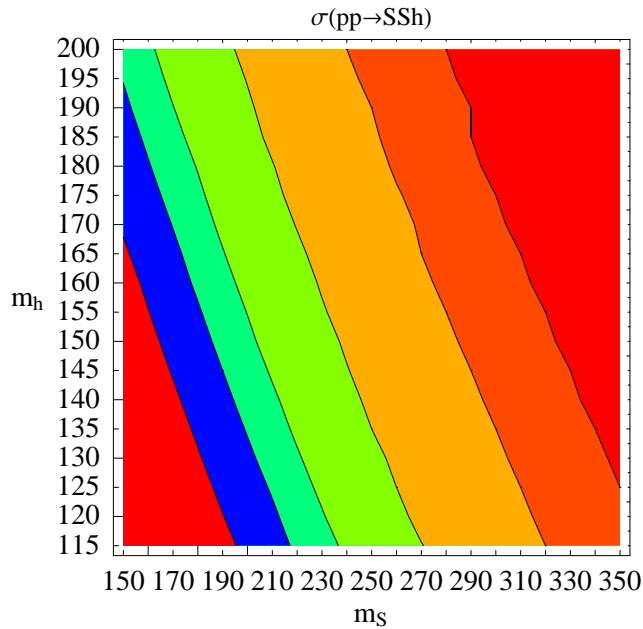}}
\caption{Production cross section at the LHC ($\sqrt{s}$=14 TeV) for the process $\sigma(pp \rightarrow SSh)$ as a function of the Higgs boson mass, $m_{h}$,  and the scalar mass $m_{S}$.    We assume a single real scalar, $S$.  For the $S-S-h$ coupling we take a value $\lambda_{hSS} = 0.75 v$, with $v=246$ GeV. From left to right contours appear at  500, 250, 150, 100, 50 ,20,10 fb.}
\end{center}
\end{figure}

We now turn to the possibility of observing the process $pp \rightarrow SSh$.  In Figure \ref{Fig:XXh} we show contours of the production cross section for this process.  We assume a single real scalar, $S$.  For the $S-S-h$ coupling we take a value $\lambda_{hSS} = 0.75 v$, with $v=246$ GeV.  Over the range shown, the cross section ranges from nearly half a picobarn down to less than 10 femtobarns.  Due to phase space suppression, the cross section is a very rapidly falling function of the $S$ mass.  Assuming no signal is found for the $S$ field at the Tevatron, any color octets will be heavy enough that cross sections for this process will not be likely to exceed 10's of fb.  Thus, searches in channels $SSh \rightarrow 6b$ (for lighter Higgs masses) and $SSh \rightarrow 4b W^{+} W^{-}$ are likely to be challenging due to large backgrounds from QCD and $t \bar{t}$ production.    If some of the $S$ fields hadronize into long-lived charged particles, then a final state including two highly-ionizing tracks along with a Higgs boson might be sufficiently distinctive to observe above any backgrounds, independent of the final state of the Higgs boson.    However, as discussed above, current bounds on such an $S$ field would require it to be heavy,
implying a small  production cross section.    

 \section{Conclusions}
Exotic colored scalars, if present at the TeV scale, will be produced at the LHC in large numbers.  Prospects for their discovery  obviously depend on the details of their decay modes.  Here we have examined the phenomenology of particles in the $(8,2)_{1/2}$ representation.   Mass splittings within this multiplet arising from its  couplings to the Higgs may allow for two-body decays of a heavier scalar to a lighter one and a $W$ or $Z$ gauge boson.    We find that these couplings must be  rather large for these two-body decays to be kinematically accessible, given  the lower-bound on their masses implied by Tevatron results.  A renormalization group analysis  then shows that these large couplings tend to hit Landau poles at fairly low scales.

Although it is certainly possible that  whatever additional new physics comes along with the octets significantly changes these RGEs, we have chosen to study the case where the octets'  decays are dominated by third-generation quarks. Our 
study of pair-production events leading to $t {\overline t} t {\overline t}$, $b {\overline b} b {\overline b}$, and  $b {\overline b} t {\overline t}$ final states suggests that the octets should be accessible at the LHC up to masses exceeding a TeV.  For these heavy octets, the top quarks produced  are quite boosted  and so their decay products are not easily resolved, but a method designed to identify top candidates by looking for composite jets with appropriate invariant masses leads to efficient top reconstruction.   
We find that observation in the single-production channel is much more difficult,  once precision electroweak constraints are taken into account. 

\section*{Acknowledgments}
The work of DP and AP is supported by the Michigan Center for Theoretical Physics.   The work of  DTS was supported by NSF grant PHY-0555421.  Thanks to Stephen Ellis, Jonathan Walsh, and James Wells for useful discussions.

\section*{Appendix: Renormalization of $\lambda_{i}$}

New scalar couplings will affect the running of the Higgs boson quartic couplings, possibly causing a Landau pole at very low energies. In principle, they might also become non-perturbative themseleves.  The RGE for the Higgs boson quartic coupling is modifed as:
 \begin{equation}
16\pi^2 \frac{d\lambda}{dt} = 24 \lambda^2 + 48 \lambda_1^2 + 16 \lambda_2^2+16\lambda_3^2 -(3g'^2+9g^2-12 y_t^2)\lambda + \frac{3}{8}g'^4 + \frac{3}{4} g'^2 g^2 +\frac{9}{8} g^4 - 6 y_t^4
\end{equation}
where $t=\ln{Q}$.  

The new scalar couplings are modified as:
\begin{eqnarray}
16\pi^2 \frac{d\lambda_1}{dt} &=& 16 \lambda_1^2 + 8 \lambda \lambda_1 - (\frac{3}{2}g^{'2}+ \frac{9}{2}g^{2} - 6 y_t^2)\lambda_1 \nonumber \\
&& - (\frac{3}{2}g^{'2}+\frac{9}{2} g^2 + 18 g_s^{2}) \lambda_1 + 6 \eta_U^2 y_t^2 \lambda_1 \nonumber \\
&& + \frac{3}{8} g^{'4} + \frac{3}{8} g^4 + \frac{3}{4}  g^{'2}g^2 \nonumber \\
&&-6 \eta_U^2 y_t^4 
\end{eqnarray}
\begin{eqnarray}
16\pi^2 \frac{d\lambda_2}{dt} &=& 8 \lambda_2^2 + 16 \lambda \lambda_2 - (\frac{3}{2}g^{'2}+ \frac{9}{2}g^{2} - 6 y_t^2)\lambda_2 \nonumber \\
&& - (\frac{3}{2}g^{'2}+\frac{9}{2} g^2 + 18 g_s^{2}) \lambda_2 + 6 \eta_U^2 y_t^2 \lambda_2 \nonumber \\
&&-6 \eta_U^2 y_t^4
\end{eqnarray}
\begin{eqnarray}
16\pi^2 \frac{d\lambda_3}{dt} &=& 8 \lambda_3^2 + 16 \lambda \lambda_3 - (\frac{3}{2}g^{'2}+ \frac{9}{2}g^{2} - 6 y_t^2)\lambda_3 \nonumber \\
&& - (\frac{3}{2}g^{'2}+\frac{9}{2} g^2 + 18 g_s^{2}) \lambda_3 + 6 \eta_U^2 y_t^2 \lambda_3 \nonumber \\
&&-6 \eta_U^2 y_t^4.
\end{eqnarray}
We have ignored the RGEs and contributions of $\lambda_4$ and higher.  Inclusion of these couplings would only strengthen the bounds coming from perturbativity considerations.


\end{document}